\documentclass[reprint,10pt,frontmatterverbose, showpacs,preprintnumbers,nofootinbib,showkeys, aps,prd,longbibliography,floatfix]{revtex4-1}

\usepackage{tensor}
\usepackage{graphicx}
\usepackage{amsmath}
\usepackage{amssymb}
\usepackage{enumerate}
\usepackage{subfigure}
\usepackage{tabularx}
\usepackage{appendix}
\usepackage{tabularx}
\usepackage{caption}
\usepackage{blindtext}
\usepackage[colorlinks=true, pdfstartview=FitV, linkcolor=blue, citecolor=red, urlcolor=black, breaklinks=true]{hyperref}
\newcommand{\be}{\begin{equation}}
\newcommand{\ee}{\end{equation}}
\newcommand{\ben}{\begin{eqnarray}}
\newcommand{\een}{\end{eqnarray}}
\newcommand{\bes}{\begin{subequations}}
\newcommand{\ees}{\end{subequations}}
\def\bal#1\eal{\begin{align}#1\end{align}}
\DeclareMathOperator{\Tr}{Tr}

\begin{document}

\title{Glueballs in the Klebanov-Strassler Theory: Pseudoscalars vs Scalars}
\author{Cornélio Rodrigues Filho}\email{cornelio@fisica.ufrn.br}
\affiliation{Departamento de F\'\i sica Teorica e Experimental, Universidade Federal do Rio Grande do Norte, Campus Universit\'ario, Lagoa Nova, Natal-RN  59078-970, Brazil}

\begin{abstract}
\begin{center}
    \textbf{Abstract}
\end{center}
We discuss the $0^{+-}$ singlet sector of glueballs in the Klebanov-Strassler theory. We report the results of a numerical study of the linearized equations in the Klebanov-Strasller background and make a comparison with the spectrum of the scalar sector. While for four towers of the total six towers of massive pseudoscalar states our results match the spectrum of the corresponding towers of scalars, the values for the remaining two towers diverge with those of the scalars. We discuss possible interpretations of this divergence.

\end{abstract}
\date{\today}
\keywords{Gauge/Gravity Duality, Klebanov-Strassler Theory, Singlet Glueballs, Pseudoscalars vs Scalars.}
\maketitle

\section{Introduction}
The example of the gauge/gravity duality \cite{Maldacena:1997re,Gubser:1998bc,Witten:1998qj} proposed by Klebanov and Strassler (KS) \cite{Klebanov:2000hb} consists of the  dual description of the  $\mathcal{N}=1$ supersymmetric non-conformal gauge theory by a solution of the type IIB supergravity equations. For reviews of this duality see \cite{Herzog:2001xk,Herzog:2002ih}. Some important features of the KS theory include the chiral symmetry breaking and an unconventional renormalization group (RG) flow \cite{Strassler:2005qs}, which among other things leads to color confinement. The confinement in this model is observed as follows: in the KS theory the $SU(M+N)\times SU(N)$ gauge group undergoes a cascading flow to a strongly coupled $\mathcal{N}=1$ supersymmetric $SU(M)$ gauge theory, which exhibits confinement. In the infrared regime one can separate the pure $\mathcal{N}=1$ Supersymmetric Yang-Mills sector within the full theory and study its low energy states (glueballs) using classical supergravity approximation. The holographic methods have a big advantage over standard field theory approach, since the latter does not have analytical access to the information about glueballs.

In the last years a series of works  \cite{Caceres:2000qe,Berg:2005pd,Berg:2006xy,Dymarsky:2006hn,Benna:2006ib,Dymarsky:2007zs,Benna:2007mb,Dymarsky:2008wd,Gordeli:2009nw,Gordeli:2013jea,Melnikov:2020cxe} have used the KS theory to estimate the behavior of masses of the glueballs for pure $\mathcal{N}=1$ theory. In these works the glueball states that are singlets under the global $SU(2)\times SU(2)$ symmetry of the KS theory \cite{Klebanov:2000hb} were considered. The holographic techniques replace the computation of the two-point correlation functions in the field theory by the analysis of the linearized equations of classical supergravity \cite{Csaki:1998qr,Brower:2000rp}. For the analysis beyond the singlet regime see \cite{Elander:2009bm,Elander:2010wd,Elander:2012yh,Elander:2014ola,Elander:2017cle,Elander:2017hyr,Elander:2018aub}.

As an alternative to holography, the spectrum of the glueballs can be estimated from lattice calculations \cite{Morningstar:1999rf,Teper:1998kw,Chen:2005mg,Lucini:2004my,Lucini:2010nv,Holligan:2019lma,Athenodorou:2020ani,Gregory:2012hu}. The predictions of the two methods can be compared.

The main goal of this paper is the discussion of the spectrum of the $0^{-+}$ (pseudoscalar) glueballs. The linearized equations and preliminary results of the spectrum calculation were recently presented in \cite{Melnikov:2020cxe}. Here we provide some details of the numerical analysis and suggest an alternative interpretation of the results to that presented in \cite{Melnikov:2020cxe}. In particular, we discuss the possibility that the results on the pseudoscalar glueballs indicate that the spectrum of the scalars known from \cite{Berg:2005pd,Berg:2006xy} must be corrected.

This letter is organized as follows. In section \ref{glue} we present a short discussion of the derivation of the pseudoscalar glueballs. We also discuss the consistency checks of the results. In section \ref{hvsl} we apply the numerical approach to estimate the spectrum of this sector and make a comparison with the spectrum of the scalar glueballs. In section \ref{conclusion} we present our  conclusions and some final remarks.

\section{Pseudoscalar Glueballs}
\label{glue}
The Klebanov-Strassler theory \cite{Klebanov:2000hb} provides a possible setup to analyze the spectrum of glueballs, which are expected to constitute the low energy spectrum of $\mathcal{N}=1$ Yang-Mills Theory. The glueballs are bound states mainly composed of gluons. Their classification is given in terms of the $J^{PC}$ quantum numbers, where $J$ is the spin, $P$ is the parity and $C$ is the charge conjugation quantum numbers. The action of these discrete symmetries in the KS theory was discussed in some detail in \cite{Melnikov:2020cxe}. Although, classical supergravity approximation can not be used to capture the glueball states with high spin, the structure of the low spin spectrum already contains some interesting information.

For our purposes we are interested in the masses of glueballs, which are singlets under the global $SU(2)\times SU(2)$ symmetry \cite{Berg:2005pd,Berg:2006xy,Dymarsky:2006hn,Benna:2006ib,Dymarsky:2007zs,Benna:2007mb,Dymarsky:2008wd,Gordeli:2009nw,Gordeli:2013jea,Melnikov:2020cxe} of the KS theory. It is these states that we expect to match the low energy limit of the pure  $\mathcal{N}=1$ SYM. Specifically, we are interested in the $0^{-+}$ glueballs that were recently described in \cite{Melnikov:2020cxe}. In this section we review the derivation of the equations for the pseudoscalars.

The KS background is a type IIB supergravity solution with a warped deformed conifold metric \cite{Candelas:1989js} of $AdS^5\times T^{11}$ topology. The metric is supported by fluxes of the $F_5$, $F_3$ and $H_3$ forms ($F_5$ is a self-dual R-R 5-form, $F_3$ is the R-R 3-form and $H_3$ is the NS-NS 3-from from the bosonic sector of the type IIB supergravity). The solution also assumes that $C=\Phi=0$, which  are the dilaton and R-R axial scalar, respectively. 

For reviews of the gravity solution of KS theory see \cite{Herzog:2001xk,Herzog:2002ih}.
The pseudoscalar sector can be described by the following fluctuations of the type IIB supergravity background \cite{Klebanov:2000hb}:

\begin{widetext}
\ben
\delta (ds^2_{T^{1,1}})&=&B(g^1\cdot g^4-g^2\cdot g^3),\label{dg13}\\
\delta (ds^2_5) &=& (*_4{d a}+A d\tau ) \wedge g^5,\label{daA}\\
\delta C&=&C, \label{C}\\
\delta C_{2}&=&C_{2}^{-}(g^{1}\wedge g^{2}-g^{3}\wedge
g^{4})+\\ \nonumber &&+C_{2}^{+}(g^{1}\wedge g^{2}+g^{3}\wedge g^{4}),\label{cpcm}\\
\delta B_{2}&=&B_{2} \left( g^{1}\wedge g^{3}+g^{2}\wedge
g^{4}\right)\label{b2},\\
\nonumber
\delta F_{5} & =& \frac{lG^{55}}{2}\Bigg\{\Big[\partial_{\mu}(a +\phi_{1})dx^{\mu}+( A+\phi _{2}) d\tau\Big] \wedge g^{1}\wedge g^{2}\wedge g^{3}\wedge
g^{4} -
 \sqrt{-G}( G^{11}G^{33})^{2}\Big[h^{1/2}  \ast_4d( a -\phi _{1}) \wedge d\tau  +\\
 &&+
(A -\phi _{2}) 
 G^{55}d^{4}x\wedge g^{5} 
\Big] - h^{1/2}\sqrt{-G} 
G^{11}G^{33}( G^{55})^{2}*_4d\phi
_{3} \wedge dg^{5}+\partial_{\mu}\phi _{3}dx^{\mu}\wedge d\tau
\wedge g^{5}\wedge dg^{5}\Bigg\}.
\label{df5}
\een

\end{widetext}
Here we consider fluctuations of the metric, R-R axial scalar, R-R 2-form potential, NS-NS 2-form potential and R-R 5-form potential, introducing a set of ten unknown functions of the coordinates $x^\mu\equiv\vec{x}=(t,x,y,z)$ and $\tau$ ($AdS_5$ radial variable),
\be
a\,,A\,,B\,,B_2\,,C_2^-\,,C_2^+\,,\phi_1\,,\phi_2\,,\phi_3.
\ee
Note that \eqref{df5} is constructed in such a way to guarantee the self-duality condition, $ F_5=*F_5$. Functions $\phi_2$ and $\phi_3$ can be eliminated through the Bianchi identity  for $F_5$.

Perturbations \eqref{dg13}-\eqref{df5} produce a coupled set of eight ordinary differential equations \eqref{eqf3}-\eqref{eqA} (seven of which are of second order and one is of first order). Function $A$ enters the equations without second derivatives. In total there are eight unknown modes for eight ODE's. 

The system of the linearized equations looks rather complicated, so a number of consistency checks were carried out to ensure their validity. The first check is the gauge invariance of the system. (See the action of the gauge transformations in equation \eqref{gauge} below.)

Another non-trivial check is imposed by supersymmetry arguments. As predicted in \cite{Gordeli:2009nw,Gordeli:2013jea}, the pseudoscalar sector has to be characterized by six physical modes. Indeed, this is confirmed by the equations. As mentioned in  \cite{Melnikov:2020cxe}, not all equations are linear independent so that it is possible to reduce the full system to a set of six equations and modes via the following arguments. First, by gauge freedom we can set $a=0$ and, as a consequence of the linear dependence, one can drop the second order ODE related to $a''$, \eqref{eqa}. Second, we use the first order ODE, \eqref{eqA}, to write the mode $A$ in terms of the remaining ones. Finally, we substitute the algebraic expression for $A$ in the remaining six equations and find the expected result. 

However, this particular system has shown to be problematic in the numerical analysis. A more appropriate choice is to set $A=0$ and work with seven second order ODE's \eqref{eqf3}-\eqref{eqa}. First order ODE \eqref{eqA} imposes a set of constraints which the remaining modes must obey. These constraints project onto the six physical modes. 

In table \ref{tab:dims} we show the physical modes together with their conformal scaling dimensions obtained from the analysis of the asymptotic expansion at $\tau\to\infty$. The fact that the dimensions are integer and match the expectations from the supersymmetry analysis (see \cite{Melnikov:2020cxe}) provides another check of the consistency of the system of equations.

\begin{table}[htb]
\begin{center}
\begin{tabular}{l|cccccc}
\hline
Mode & $\phi_3$ & $P_2$ & $Q_2$ & $C_2^+$ & $C$ & $B$  \\ \hline

Dimension, $\Delta$ & $5$ & $3$ & $7$ & $4$ & $4$ & $3$    \\ 
\hline
\end{tabular}
\end{center}
\caption{Physical modes and the corresponding scaling dimensions of the dual operators of the pseudoscalar sector. Here $P_2=B_2-C_2^-$ and $Q_2=B_2+C_2^-$ .}
\label{tab:dims}
\end{table}

The last consistency check will be discussed in the next section, when we are going to use numerical calculations to estimate the spectrum of the pseudoscalar glueballs. In addition we will compare our results with those ones for the scalar sector \cite{Berg:2006xy}.

\section{Numerical Results}\label{hvsl}
Let us begin this section explaining the numerical approach employed to solve the system of equations described in the last section (also see equations \eqref{eqf3}-\eqref{eqA} in the appendix). A common method to solve the eigenvalue problem numerically is the shooting method, which involves replacing the boundary problem with the initial value one on either of the boundaries of the system $\tau\to 0$ (IR), or  $\tau\to \infty$ (UV). However, this technique can not be applied directly to coupled systems of equations. Instead, one uses the generalization that is called Midpoint Determinant Method (MDM) \cite{Berg:2006xy}, which combines both the IR and UV initial problems. In the MDM one builds a $2n\times 2n$ quadratic matrix for $n$ fields from the UV and IR numerical solutions for the fields and their derivatives at some intermediate point $\tau_{mid}$, which is a point where the solutions are matched, i. e.,
\be\label{deta}
\gamma=
\left(
\begin{array}{cc}
x_{IR} & x_{UV}\\
\partial x_{IR} & \partial x_{UV}
\end{array}\right)
_{\tau =  \tau_{mid}}.
\ee
Here $x_i$ is the set of fields, $i=1,\ldots,n$, and $\partial x_i$ is the set of derivatives. The eigenvalues are given by the condition $\det\gamma=0$, so that $\gamma$ is a function of $\tilde{m}^2$. To be precise as $\det\gamma$ is an oscillatory function of $\tilde{m}^2$ one finds the eigenvalues looking for the loci, where the determinant changes of sign. 

To run the MDM it is necessary to specify five parameters, they are $\tau_{IR}$ and $\tau_{UV}$ the IR and UV cut-offs respectively, middle point $\tau_{mid}$, the step $\delta \tau$ and $n_{imposed}$. The latter as explained in \cite{Berg:2006xy} is used to control the behavior of the numerical solutions when some modes fluctuate dramatically faster than the other ones. After that, one constructs the analytic solutions in the IR and UV ends to be fed as initial conditions. 

As we have mentioned the pseudoscalar sector is described by six physical modes and a convenient way to express this sector is through seven second ODE's where we have to add a nonphysical mode. Even in the presence of this mode we expect to separate the six physical families of eigenvalues. 

We now move on to discuss the results obtained by application of the MDM procedure. In our analysis we calculate a $14\times 14$ matrix with the following parameters: $\tau_{IR}=0.1\,,\tau_{UV}=20\,,\tau_{mid}=1\,,\delta \tau=0.01\,,n_{imposed}=500$. We use the boundary conditions obtained in \cite{Melnikov:2020cxe}. In order to compare our results with those found in \cite{Berg:2006xy} we adopt their mass normalization, i.e,  $m^2=0.9409\tilde{m}^2$. 

Before we present the results of the numerical analysis we would like to mention that we also investigated the $12\times 12$ matrix scenario, with only six physical modes. In this case we computed the $12\times 12$ matrix dropping different modes to build it, but this approach does not provide consistent results. In particular, it does not capture the expected six towers of eigenvalues. For $14\times 14$ case, 
the results offer a good consistency, independent from the variation of the parameters.

\begin{figure}[t]
    \centering
      \includegraphics[width=0.47\textwidth]{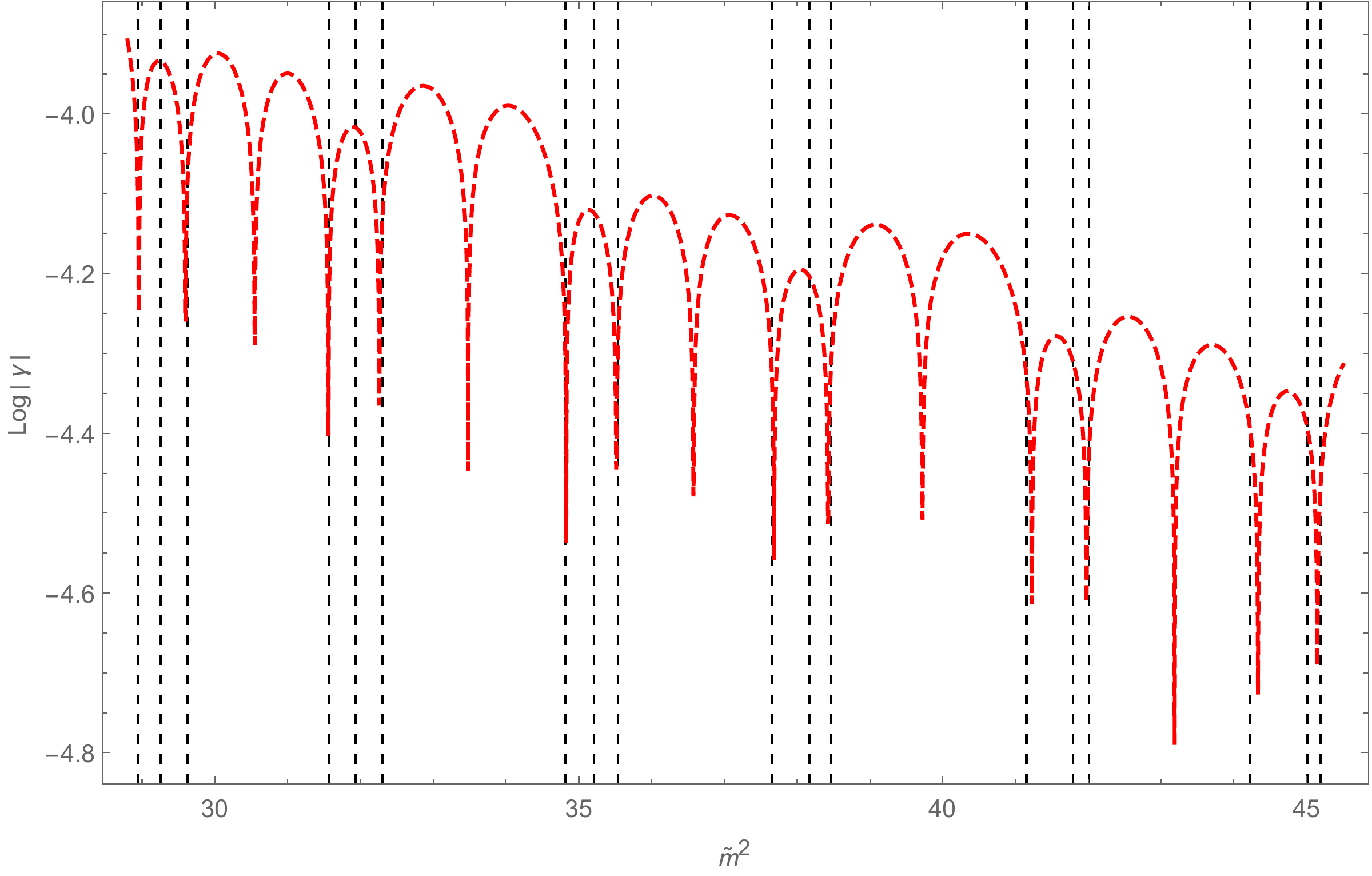} 
    \caption{The heavier values of $m^2$. Red line corresponds to our values for MDM analysis and the dashed lines are the values of \cite{Berg:2006xy}.}\label{fig:my_label}
\end{figure}

Finally, we provide the results found for the MDM analysis. In table \ref{tab:mid7} we show the lower values of $m^2$, in table \ref{tab:full} we summarize the first 65 values of $m^2$ and in addition we also include the results from the scalar sector computed in \cite{Berg:2006xy} to perform a comparison between the results of both sectors latter. The heavier part of table \ref{tab:full} is shown in figure \ref{fig:my_label}, our results are in red and the dots in vertical are the results of \cite{Berg:2006xy}. 

\begin{table}[b]
\begin{center}
 \begin{tabular}[c]{|l|l|l|l|l|l|l|l|l|}
       \hline 
\bf{$n$}                 & \bf{1}      & \bf{2}      & \bf{3}     & \bf{4}    & \bf{5}    & \bf{6}     \\\hline 
$m^2$ &  0.273    &  0.513 & { 0.946} & { 1.38} & 1.67  & { 2.09} \\\hline
\bf{$n$}  & \bf{7}    & \bf{8}    & \bf{9} &\bf{10} &\bf{11} &\bf{12}     \\ \hline
${m}^2$ & 2.34  & { 2.73} & 3.33  & { 3.63}  & 4.24 & 4.43   \\\hline
\bf{$n$}     & \bf{13}     & \bf{14}     & \bf{15}    & \bf{16}    & \bf{17}    & \bf{18}   \\ \hline
$m^2$   & { 4.96}   & 5.44  & 5.63  &  { 6.25} &   6.63  &    {6.96} \\ \hline
      \end{tabular}
      \captionof{table}{The lowest values of $m^2 $ found by MDM analysis.}\label{tab:mid7} 
      \end{center}
\end{table}
    
From figure \ref{fig:my_label} we can extract some interesting results. First, by supersymmetry arguments the  $0^{-+}$ and $0^{++}$ sectors must share the same spectrum. In \cite{Berg:2006xy} seven families of eigenvalues were obtained, but one of these towers of eigenvalues was found in \cite{Gordeli:2009nw} to be a superpartner of a $1^{++}$ glueball. Then, as can be seen in figure \ref{fig:my_label} there is a periodicity among the eigenvalues so that as expected we are able to separate the spectrum of the pseudoscalar glueballs into six towers of eigenvalues in such a way that it gives us another consistency test. However, we also see that only four towers have good matches with the results of \cite{Berg:2006xy}. The remaining two towers show a very distinct behavior from the $0^{++}$ values. So, this apparent tension between the $0^{-+}$ and $0^{++}$ spectra brings up some questions.

The first question is about the validity of the linearized equations. We believe that their derivation is correct and we showed that they pass a number of consistency checks. It is more natural to expect that the problem is in the numerical analysis. In our case we also used the same MDM method as in \cite{Berg:2006xy} to compute the spectrum. The results of $0^{++}$ were independently checked in \cite{Elander:2014ola}. Therefore, in principle we are led to conclude that the $0^{++}$ spectrum is the correct one. On the other hand, we have not been able to improve our numerical analysis in such a controllable way that the convergence to the spectrum of~\cite{Berg:2006xy} is improved for the two inconsistent modes. In particular, table \ref{tab:check} demonstrates that the corresponding eigenvalues, like the value $m^2=17.68$, are rather stable with respect to the variation of the parameters of the numerical analysis. 

Let us give some additional considerations in support of our spectrum. First, we note that our values give better quadratic fits for the low eigenvalues, as compared to the eigenvalues of \cite{Berg:2006xy} (for other glueballs in the KS background the quadratic fit usually works quite well),
\ben
m^2&\approx&0.269 n^2+1.041 n+1.062\,, \label{fit1}\\
m^2&\approx&0.273 n^2-0.148 n+0.172\,,\\
m^2&\approx& 0.270 n^2 + 0.250 n +0.115\,,\\
m^2&\approx&0.268 n^2+0.775 n+0.233\,,\\
m^2&\approx&0.272 n^2+1.693 n+2.46\,,\\
m^2&\approx&0.272 n^2+1.975 n+3.383. \label{fit6}
\een
These fits match well even the lowest states of table \ref{tab:full}, which is not the case of the fits obtained in~\cite{Berg:2006xy}. To illustrate this we plot in figure \ref{fig:fits} our fits and those of \cite{Berg:2006xy}. Note that the fits of \cite{Berg:2006xy} do not capture several values of $m^2$. Besides that, there are several degenerate states as seen in figure \ref{fig:fits} B. Our fits, on the other hand, exhibit a good match with the values of $m^2$ in table \ref{tab:full}. We also do not observe degenerate eigenvalues.

Yet another interesting consideration comes from the comparison with the lattice results.  
The lowest states of $0^{-+}$ glueball in table \ref{tab:mid7} better matches the position of this glueball in the lattice calculations  \cite{Morningstar:1999rf,Teper:1998kw,Chen:2005mg,Lucini:2004my,Lucini:2010nv,Holligan:2019lma,Athenodorou:2020ani,Gregory:2012hu}. To be precise, an exceptional match occurs when we use the lattice results for $SU(\infty)$ extrapolation \cite{Lucini:2004my}, which is the regime where we expect the supergravity approximation to be valid. In this case our result captures the fundamental state and its excitation, see table \ref{tab:lattice}.

\begin{table}[t]
\begin{center}
\begin{tabular}{|c|c|c|c|c|c|c|c|c|}
\hline
$\tau_{IR}$ & $\tau_{UV}$ & $\tau_{mid}$& $\delta \tau$ & $n_{imposed}$& $m^2$   \\ \hline
0.005 & 15 & 1 & 0.01& 500 & 17.6838\\ \hline
0.005 & 20 & 1 & 0.001& 500 & 17.6861 \\ \hline
0.01 & 22 & 4 & 0.01& 400 & 17.6858 \\ \hline
0.1 & 20 & 1 & 0.01& 500 & 17.6782 \\ \hline
0.1 & 20 & 1 & 0.001& 500 & 17.6810 \\ \hline
0.2 & 18 & 2 & 0.01& 400 & 17.6709 \\ \hline
\end{tabular}
\end{center}
\caption{Values of $m^2 \approx 17.68$. for different chooses of the parameters of MDM.} \label{tab:check}
\end{table}

\begin{table}[t]
\begin{center}
\begin{tabular}{l|cccccc}
\hline
Ratio & \cite{Lucini:2004my} & This Work & \cite{Berg:2006xy}   \\ \hline
$m_{0^{++}}/m_{2^{++}}$ & $0.689$ & $0.700$ & $0.640$     \\ \hline
$m_{0^{++*}}/m_{2^{++}}$ &  $1.264$   &     $1.264$& $1.249$  \\
\hline
$m_{\lambda\lambda}/m_{2^{++}}$ &  --   &     $0.511$ & $0.421$  \\
\hline
$m_{\lambda\lambda^\ast}/m_{2^{++}}$ &  --   &     $0.952$ & $0.894$  \\
\hline
\end{tabular}
\end{center}
\caption{Ratio of masses of the glueballs. As usual we normalize the spectrum by $m_{2^{++}}$. For holographic results we use the $m_{2^{++}}$ of \cite{Berg:2006xy}.} 
\label{tab:lattice}
\end{table}

\section{Conclusion and Final Remarks} 
\label{conclusion}
In this paper we have discussed the pseudoscalar glueballs of the KS theory. We have explained the derivation of the linearized supergravity equations that compute the spectrum and reviewed several consistency checks. In the numerical analysis it was possible to identify six families of eigenvalues. Four of those have good matches with the spectrum of scalar glueballs. The fact that the two remaining towers do not match with the remaining scalars indicates that results of the numerical approach in either $0^{-+}$ or $0^{++}$ sectors are incorrect. Here we have argued that the results of our numerical analysis are consistent. We also observed that our results show more expected behavior than those in the scalar sector, as demonstrated by fits \eqref{fit1}-\eqref{fit6}, figure~\ref{fig:my_label} and table \ref{tab:lattice}. In particular, our fits allow to reliably disentangle all six modes. 

Another interesting conclusion that can be made based on our spectrum is the position of the lightest superpartner states in the $\mathcal{N}=1$ SYM theory, as compared to the lattice predictions of the purely bosonic sector. The separation of the six towers allows to conclude that the lightest state is likely to be the fermion bilinear $\lambda\lambda$. Its mass ratio with the $2^{++}$ glueball, shown in table~\ref{tab:lattice}, shows that it is unlikely the scalar $\Tr F^{\mu\nu}F_{\mu\nu}$. Moreover, the next excited state in the same tower, which corresponds to the value $m_{\lambda\lambda^{*}}/m_{2^{++}}=0.952$, does not have a good match in the bosonic sector as well, so for the six lightest states we find the hierarchy
\be\nonumber
m_{\lambda\lambda}<m_{0^{++}}< m_{2^{++}} < m_{0^{-+}}<m_{\lambda\lambda^\ast} < m_{0^{++\ast}}
\ee
Finally, we note that the $0^{++}$ and $0^{-+}$ sectors were derived in a different way. In \cite{Berg:2006xy} the derivation was done through a sigma model approach and in  \cite{Melnikov:2020cxe} as well as in  \cite{Dymarsky:2006hn,Benna:2006ib,Dymarsky:2007zs,Benna:2007mb,Dymarsky:2008wd,Gordeli:2009nw} it was done through the type IIB supergravity derivation. In a future work we plan to check the sigma model equations in the $0^{++}$ sector through the direct linearization of the supergravity equations and derive the sigma model equations for the pseudoscalars. We hope that this will allow us to settle the divergence issue.

\acknowledgements{The author would like to thank D. Melnikov for collaboration and useful comments on the manuscript of this letter. I would also like to thank Daniel Elander and Maurizio Piai for useful correspondence. I thank the financial support by Brazilian Ministry of Education. This research was developed within the project of the Brazilian agency CNPq, process 433935/2018-9.}

\appendix
\section{Linearized Equations and Gauge Invariance}
\label{eqs}
In this appendix we show the linearized equations that are a result of the fluctuations  \eqref{dg13}-\eqref{df5}.
\begin{widetext}
\begin{scriptsize}
\begin{multline}
  \left( \frac{27IK^{4}\sinh ^{2}\tau }{32}\left( \frac{I^{\prime
}K^{6}\sinh ^{2}\tau }{I^{3/2}}\phi _{3}\right) ^{\prime }\right) ^{\prime }+%
\frac{3I^{\prime }K^{4}\sinh ^{2}\tau \phi _{3}}{4I^{1/2}}\left( \frac{%
9IK^{4}\sinh ^{2}\tau }{8}\tilde{m}^{2}-1\right)-\left( \frac{3I^{\prime}K^{4}\sinh ^{2}\tau }{4I^{1/2}}a\right) ^{\prime }+
\frac{3I^{\prime }K^{4}\sinh ^{2}\tau A}{4I^{1/2}}+\\+\frac{2^{1/3}I^{\prime
}B_{2}}{K}+\left( \frac{2^{1/3}I^{\prime }}{K}\right) ^{\prime
}C_{2}^{-}
+\left( \frac{2^{1/3}I^{\prime }\cosh \tau }{K}\right) ^{\prime}C_{2}^{+}=0;   
\label{eqf3}
 \end{multline}
 
\begin{multline}
\left( \frac{\cosh ^{2}\tau +1}{I\sinh ^{2}\tau }C_{2}^{-\prime }\right)
^{\prime }-\frac{C_{2}^{-}}{I}+\frac{\left( \cosh ^{2}\tau +1\right) \tilde{m%
}^{2}C_{2}^{-}}{K^{2}\sinh ^{2}\tau }+\left( \frac{2\cosh \tau }{I\sinh
^{2}\tau }C_{2}^{+\prime }\right) ^{\prime }+\frac{2\cosh \tau \tilde{m}%
^{2}C_{2}^{+}}{K^{2}\sinh ^{2}\tau }
-\left( \frac{2^{1/3}I^{\prime }B}{%
2I^{3/2}K^{2}\sinh ^{2}\tau }\right) ^{\prime }+  \frac{K^{2}B}{2^{1/3}I^{3/2}}-\\-\frac{\tau }{2\sinh \tau }\left( \frac{C}{I}%
\right) ^{\prime }+\frac{I^{\prime }B_{2}}{I^{2}}+\frac{2^{1/3}3}{8}\left( \frac{K^{2}}{I^{3/2}}\left( \frac{%
I^{\prime }}{K}\right) ^{\prime }A\right) ^{\prime } +\frac{2^{1/3}3I^{\prime }KA%
}{8I^{3/2}}+\left( \frac{I^{\prime }%
}{K}\right) ^{\prime }\frac{3\tilde{m}^{2}a}{2^{2/3}4I^{1/2}}+ \frac{27I^{\prime }K^{6}\sinh ^{2}\tau \tilde{m}^{2}\phi _{3}}{%
2^{2/3}32I^{3/2}}\left( \frac{I^{\prime }}{K}\right) ^{\prime }=0;
\label{eqC2p}
\end{multline}

\begin{multline}
\label{eqC2m}
\left( \frac{\cosh ^{2}\tau +1}{I\sinh ^{2}\tau }C_{2}^{+\prime }\right)
^{\prime }+\frac{\left( \cosh ^{2}\tau +1\right) \tilde{m}^{2}C_{2}^{+}}{%
K^{2}\sinh ^{2}\tau }+\left( \frac{2\cosh \tau C_{2}^{-\prime }}{I\sinh
^{2}\tau }\right) ^{\prime }+\frac{2\cosh \tau \tilde{m}^{2}C_{2}^{-}}{%
K^{2}\sinh ^{2}\tau }
-\left( \frac{C}{2I}\right) ^{\prime }+\frac{2^{1/3}3}{8}\left( \frac{K^{2}%
}{I^{3/2}}\left( \frac{I^{\prime }}{K}\cosh \tau \right) ^{\prime }A\right)
^{\prime }+\\+\left( \frac{I^{\prime }}{K}\cosh \tau \right) ^{\prime }\frac{3%
\tilde{m}^{2}a}{2^{2/3}4I^{1/2}}-\left( \frac{2^{1/3}I^{\prime }\cosh \tau B}{%
2I^{3/2}K^{2}\sinh ^{2}\tau }\right) ^{\prime } +\frac{27I^{\prime }K^{6}\sinh ^{2}\tau 
\tilde{m}^{2}\phi _{3}}{2^{2/3}32I^{3/2}}\left( \frac{I^{\prime }}{K}\cosh
\tau \right) ^{\prime }=0;
\end{multline}

\begin{multline}\label{eqb2}
B_{2}^{\prime \prime }-\frac{I^{\prime }B_{2}^{^{\prime }}}{I}-\frac{%
\left( \cosh ^{2}\tau +1\right) B_{2}}{\sinh ^{2}\tau }+\frac{\tilde{m}%
^{2}IB_{2}}{K^{2}}+\frac{I\sqrt{K^{3}\sinh \tau }}{2^{1/3}}\left( \sqrt{\frac{K}{I^{3}\sinh \tau }}B\right) ^{\prime }+ \frac{3I^{1/2}I^{\prime }\tilde{m}^{2}a}{2^{8/3}K}
+\frac{3I^{\prime }I\sinh
^{2}\tau }{2^{8/3}K}\left( \frac{K^{2}}{I^{3/2}\sinh ^{2}\tau }A\right)
^{\prime }+\\+\frac{2^{2/3}I^{\prime
}I}{4K}\left( \frac{C}{I}\right) ^{\prime }+\frac{I^{\prime }C_{2}^{-}}{I}
+\frac{2^{1/3}27I^{\prime 2}K^{5}\sinh ^{2}\tau \tilde{m}^{2}\phi _{3}}{%
64I^{1/2}}=0;
\end{multline}

\begin{multline}\label{eqc}
C^{\prime \prime }+\frac{2\left( K\sinh \tau \right) ^{\prime }C^{\prime }%
}{K\sinh \tau }+\frac{\tilde{m}^{2}IC}{K^{2}}+\left( \frac{I^{\prime \prime }}{I}+2\frac{I^{\prime }}{I}\frac{\left( K\sinh \tau \right) ^{\prime }}{K\sinh \tau }\right) C-\frac{2^{5/3}I^{\prime }B}{I^{3/2}K\sinh ^{2}\tau }-
\frac{3K^{6}A}{2I^{1/2}}+\frac{2^{4/3}\tau C_{2}^{-\prime }}{IK^{2}\sinh
^{3}\tau }
+\frac{2I^{\prime }C_{2}^{-}}{IK^{3}\sinh ^{3}\tau }+\frac{%
2^{4/3}C_{2}^{+\prime }}{IK^{2}\sinh ^{2}\tau }-\\-\frac{2}{IK^{2}\sinh
^{2}\tau }\left( \frac{I^{\prime }}{K}B_{2}\right) ^{\prime }=0;
\end{multline}
\begin{multline}
B^{\prime \prime }-\frac{I^{\prime }B^{\prime }}{I}+\frac{\tilde{m}^{2}IB}{%
K^{2}}+\frac{4B}{9K^{6}\sinh ^{2}\tau }+\frac{3B}{4}\left( \frac{I^{\prime }%
}{I}\right) ^{2}+\frac{K^{\prime 2}B}{K^{2}}+\frac{\left( K^{2}\coth \tau
\right) ^{\prime }B}{K^{2}}+\frac{2^{4/3}I^{\prime }\cosh \tau C_{2}^{+\prime }}{I^{1/2}K^{2}\sinh
^{2}\tau } + \frac{3IK\tilde{m}^{2}a}{2}+\frac{3I^{5/2}}{2K}\left( \frac{K^{4}}{I^{5/2}
}A\right) ^{\prime }+\\+\left( \frac{I^{\prime \prime }}{2I}-\frac{%
2^{1/3}K^{4}}{I}+2\frac{I^{\prime }}{I}\frac{\left( K\sinh \tau \right)
^{\prime }}{K\sinh \tau }\right) B -\frac{2^{4/3}I^{\prime }K}{I^{1/2}}C-2^{5/3}\sqrt{\frac{%
K}{I\sinh \tau }}\left( \sqrt{K^{3}\sinh \tau }B_{2}\right) ^{\prime }+\frac{%
2^{5/3}K^{2}C_{2}^{-}}{I^{1/2}}+\frac{2^{4/3}I^{\prime }C_{2}^{-\prime }}{%
I^{1/2}K^{2}\sinh ^{2}\tau }=0;
\end{multline}
\begin{multline}\label{eqa}
a^{\prime \prime }+\frac{2\left( K^{2}\sinh \tau \right) ^{\prime }a}{%
K^{2}\sinh \tau }-\frac{8a}{9K^{6}\sinh ^{2}\tau }+\left( \frac{I^{\prime
\prime }}{2I}+\frac{I^{\prime }}{I}\frac{\left( K^{2}\sinh \tau \right)
^{\prime }}{K^{2}\sinh \tau }+\frac{1}{4}\left( \frac{I^{\prime }}{I}\right)
^{2}\right) a-\frac{4B}{3K^{3}\sinh ^{2}\tau }-\frac{\left(
I^{1/2}K^{2}\sinh ^{2}\tau A\right) ^{\prime }}{I^{1/2}K^{2}\sinh ^{2}\tau }-\\-\left( \frac{I^{\prime }}{K}\right) ^{\prime }\frac{2^{7/3}C_{2}^{-}}{%
3I^{1/2}K^{4}\sinh ^{2}\tau }-\left( \frac{I^{\prime }}{K}\cosh \tau \right)
^{\prime }\frac{2^{7/3}C_{2}^{+}}{3I^{1/2}K^{4}\sinh ^{2}\tau }-\frac{%
2^{7/3}I^{\prime }B_{2}}{3I^{1/2}K^{5}\sinh ^{2}\tau }-\frac{9I^{\prime }}{%
8I^{1/2}}\left( \frac{I^{\prime }K^{6}\sinh ^{2}\tau }{I^{3/2}}\phi
_{3}\right) ^{\prime }=0;
\end{multline}
\begin{multline}\label{eqA}
-I\tilde{m}^{2}A+K^{2}\left( -\frac{I^{\prime \prime }}{I}+\frac{2}{\sinh
^{2}\tau }-2\frac{I^{\prime }}{I}\frac{\left( K\sinh \tau \right) ^{\prime }%
}{K\sinh \tau }\right) A+2I\left( \frac{K^{\prime }}{K}-\frac{1}{4}\frac{%
I^{\prime }}{I}\right) \tilde{m}^{2}a+I\tilde{m}^{2}a^{\prime }+\frac{4}{3I^{3/2}\sinh ^{2}\tau }\left( \frac{I^{3/2}B}{K}\right) ^{\prime }+ 
\frac{2^{5/3}K^{4}C}{3I^{1/2}}-\\-\frac{2^{7/3}I^{\prime }C_{2}^{-}}{%
3I^{1/2}K^{3}\sinh ^{2}\tau }+\frac{2^{7/3}C_{2}^{-\prime }}{%
3I^{1/2}K^{2}\sinh ^{2}\tau }\left( \frac{I^{\prime }}{K}\right) ^{\prime }+
\frac{2^{7/3}C_{2}^{+\prime }}{3I^{1/2}K^{2}\sinh ^{2}\tau }\left( \frac{I^{\prime }}{K}\cosh \tau \right) ^{\prime }+\frac{2^{7/3}}{3I^{1/2}K^{2}\sinh ^{4}\tau }\left( \frac{I^{\prime }\sinh
^{2}\tau }{K}B_{2}\right) ^{\prime }-\\-\frac{9I^{\prime 2}K^{6}\sinh ^{2}\tau 
\tilde{m}^{2}\phi _{3}}{8I}=0.  
\end{multline}
\end{scriptsize}
The gauge invariant transformations that provides a symmetry in the equations above are the following:
\begin{equation}\label{gauge}
\begin{array}{lllll}
\delta C  = 0, && \delta a  = 2G_{55}\alpha,  && \delta B=-\frac{1}{2}\varepsilon ^{4/3}h^{1/2}K%
\alpha, \\
&&&&\\
\delta C_{2}^{+}=\frac{%
\alpha}{2},  && \delta A=2G_{55}\alpha^{\prime },  &&  \delta C_{2}^{-}=\frac{1}{2}\left( 2F-1\right) \alpha, \\
&&&&\\
\delta\phi
_{3}=0,  && \delta B_{2} =\frac{\left( k-f\right) }{2}\alpha,  &&
\end{array}
\end{equation}%
where $\alpha\equiv\alpha(\vec{x},\tau)$.
\end{widetext}

\section{Numerics}
In this appendix we show the first 65 values of $m^2$ from MDM analysis in table \ref{tab:full}. We also include the $0^{++}$ glueballs results \cite{Berg:2006xy}. 
\begin{table}[h!]
\begin{center}
\resizebox{8.5cm}{!}{
\begin{tabular}{|l|l|l|l|l|l|l|l|l|l|l|l|l|l|l|l|l|}
\hline 
\bf{$n$}                 & \bf{1}      & \bf{2}      & \bf{3}     & \bf{4}    & \bf{5}    & \bf{6}    & \bf{7}    & \bf{8}    & \bf{9} &\bf{10} &\bf{11} \\\hline 
$m^2$ & {0.273}    & {0.513} & {0.946} & {1.38} & 1.67  & {2.09} & 2.34  & {2.73} & 3.33  & {3.63}  & 4.24\\\hline
${m}_{BHM}^2$ & 0.185  & 0.428  & 0.835 & 1.28 & 1.63 & 1.94 & 2.34 & 2.61 & 3.32  & 3.54  & 4.18 \\\hline\hline
\bf{$n$}                 & \bf{12}     & \bf{13}     & \bf{14}     & \bf{15}    & \bf{16}    & \bf{17}    & \bf{18}    & \bf{19}    & \bf{20}  &\bf{21} &\bf{22}    \\ \hline
${m}^2$ & 4.43    & { 4.96}   & 5.44  & 5.63  &  { 6.25} &   6.63  &    {6.96} & 7.61  & 8.09     &{8.43} & 8.93 \\\hline
${m}_{BHM}^2$ & 4.43  & 4.43  & 5.36  & 5.63  & 5.63    & 6.59    & 6.77   & 7.14   & 8.08    & 8.25 & 8.57  \\ \hline\hline
\bf{$n$}                  & \bf{23}      & \bf{24}      & \bf{25}      & \bf{26}     & \bf{27}  & \bf{28}     & \bf{29}  & \bf{30}    & \bf{31}      & \bf{32}  & \bf{33}      \\ \hline
${m}^2$   & 9.56 &{9.93} & 10.86  & 11.33 & {11.74}  &  12.51 &  13.01 & {13.63} & 14.55  &  15.09 & {15.64}  \\\hline
${m}_{BHM}^2$ & 9.54   &   9.62& 10.40  & 11.32 & 11.38  & 12.09  & 12.99  & 13.02   & 14.23 & 15.03  & 15.09  \\ \hline\hline
$n$          & \bf{34} & \bf{35} & \bf{36}  & \bf{37}    & \bf{38}      & \bf{39} & \bf{40}& \bf{41}& \bf{42}& \bf{43}& \bf{44}  \\ \hline
${m}^2$    & 16.40     & 17.0 & {17.68} & 18.84    & 19.38 & {20.06}  & 20.95   & 21.55  &{22.47}   &23.62  & 24.22 \\\hline
${m}_{BHM}^2$   & 16.19  & 16.89  & 17.03 &  18.61  &  19.22 & 19.40   & 20.79  & 21.58 &22.10   & 23.53 & 23.95      \\ \hline\hline
$n$     
 & \bf{45}  & \bf{46}& \bf{47}& \bf{48}& \bf{49} & \bf{50}  & \bf{51} & \bf{52}    & \bf{53}      & \bf{54} & \bf{55}    \\ \hline
${m}^2$  & {25.02}  & 25.97     & 26.62 & {27.62}  & 28.95  & 29.59 & {30.55} & 31.56  & 32.26 & {33.48}&34.83     \\\hline
${m}_{BHM}^2$& 24.24 & 25.94& 26.32  & 26.67  & 28.95  & 29.25  & 29.62  & 31.57 &31.93 & 32.30  & 34.82  \\ \hline
$n$ & \bf{56}& \bf{57}& \bf{58}& \bf{59} & \bf{60}&\bf{61}& \bf{62} & \bf{63}& \bf{64}&\bf{65}& \bf{66}  \\ \hline
${m}^2$     &   35.51 &{36.58} & 37.68   & 38.42  &{39.72}  & 41.22   & 41.97&{43.19} & 44.33    & 45.15&---\\\hline
${m}_{BHM}^2$  & 35.21 & 35.54  & 37.65 & 38.17      & 38.47  & 41.15 & 41.79 &  42.01  & 44.22 & 45.01 & 45.19   \\ \hline

\end{tabular}%
}
\end{center}
\caption{The values of $m^2$ computed from  MDM analysis with those ones of \cite{Berg:2006xy}.} 
\label{tab:full}
\end{table}

\begin{figure}[b]
    \centering
    \includegraphics[width=0.48\linewidth]{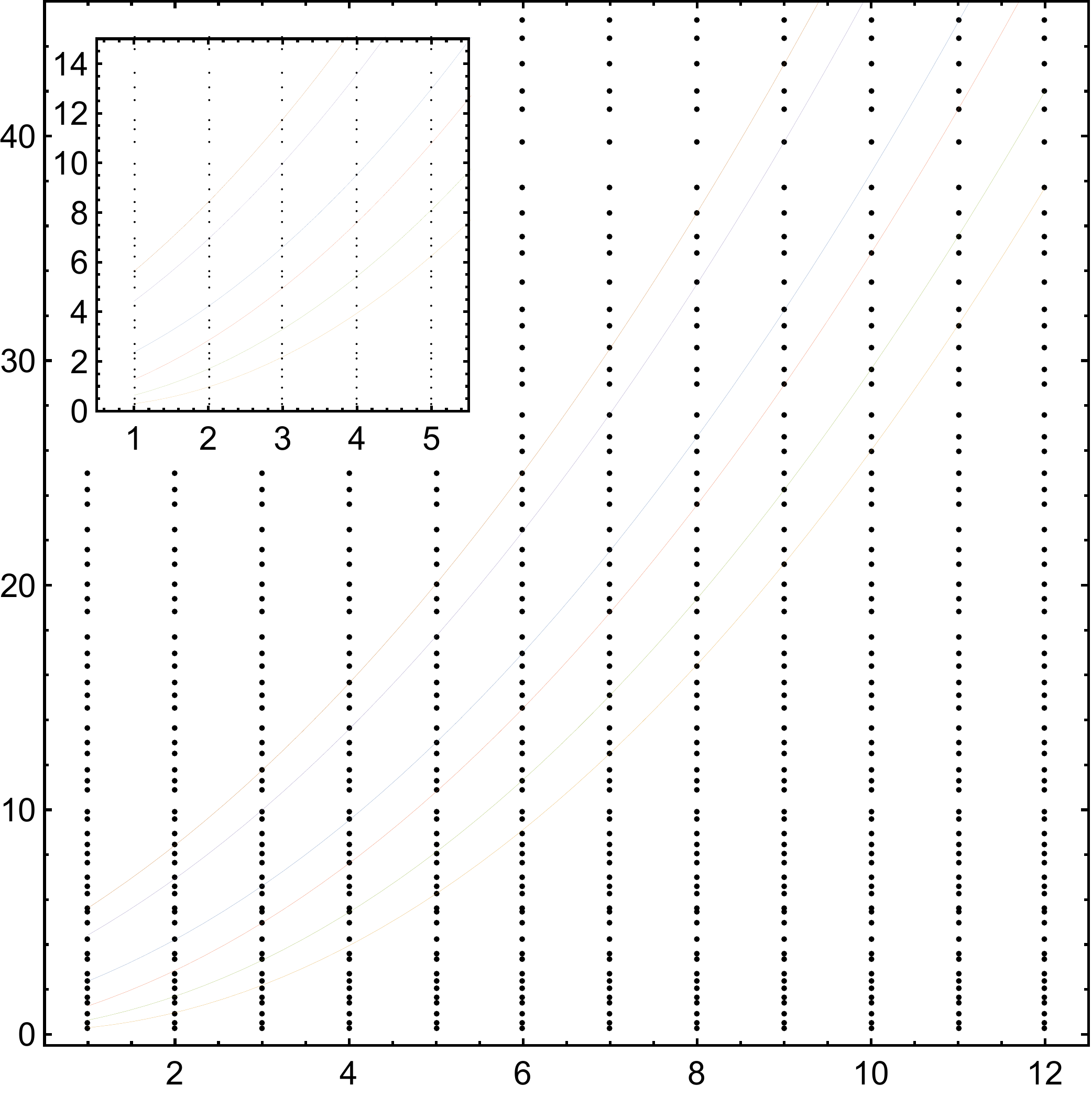}
    \hfill
    \includegraphics[width=0.48\linewidth]{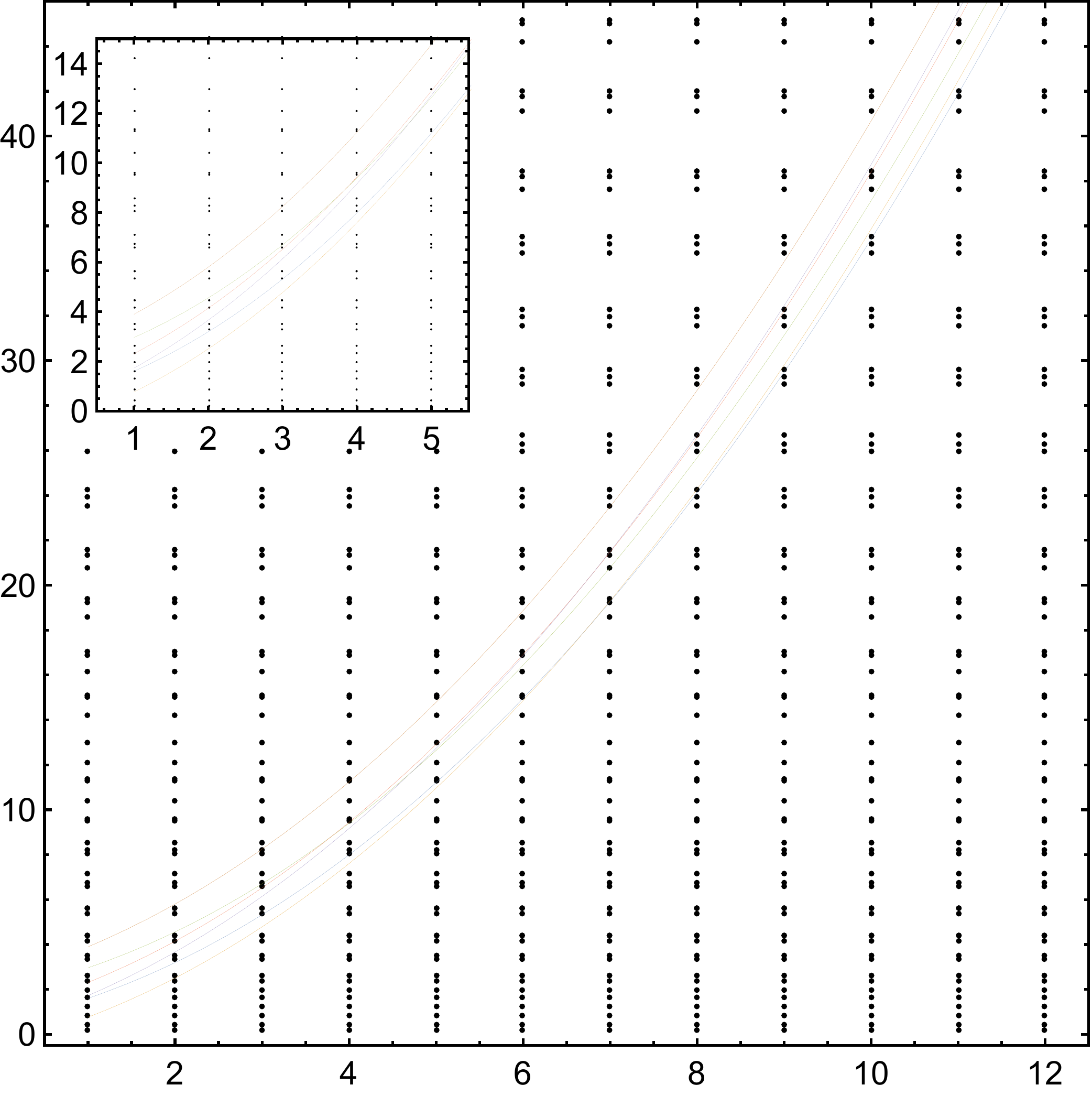}
    \caption{The lines correspond to quadratic fits and the vertical dots are the periodically repited values values of the MDM analyses for our case as well as of \cite{Berg:2006xy}. Our values are shown in the left panel and those ones of \cite{Berg:2006xy} in the right one. The insets show the lowest values of $m^2$.}
    \label{fig:fits}
\end{figure}



\bibliography{references}

\end{document}